\documentclass{article}
\usepackage{graphicx}

\font\tbf=cmbx10 scaled \magstep2
\font\tenmib=cmmib10 \textfont"E=\tenmib
\font\tenbsy=cmbsy10 \textfont"F=\tenbsy

\def\be{\begin{equation}}
\def\ee{\end{equation}}
\def\bea{\begin{eqnarray}}
\def\eea{\end{eqnarray}}

\title{\large \tbf A short answer to critics
\\ \vskip 0.02truecm of our article ``Eppur si espande''}

\author{Marek A. Abramowicz$^{1,2}$, Stanis{\l}aw Bajtlik$^{2}$,\\
Jean-Pierre Lasota$^{3,4}$ \& Audrey Moudens$^{5}$\\
{\small $^{1}$Physics Department, G{\"o}teborg University, SE-412-96 G{\"o}teborg, Sweden}\\
{\small $^{2}$Nicolaus Copernicus Astronomical Center, Bartycka 18, 00-716
Warszawa, Poland}\\
{\small $^{3}$Institut d'Astrophysique de Paris, UMR 7095 CNRS, Universit\'e
P. et M. Curie},\\
{\small 98bis Bd Arago, 75014 Paris, France}\\
{\small $^{4}$Astronomical Observatory, Jagiellonian University,}\\
{\small ul. Orla 171, 30-244 Krak\'ow, Poland}\\
{\small $^{5}$ Institut de Physique, Universit{\'e} de Rennes, 35042 Rennes, France}\\
}

\begin{document}

\maketitle

\begin{abstract}
Recently we presented a formal mathematical proof that,
contrary to a widespread misconception, cosmological expansion
{\it cannot} be understood as the motion of galaxies in
non-expanding space. We showed that the cosmological redshift
{\it must be} physically interpreted as the expansion of space.
Although our proof was generally accepted, a few authors
disagreed. We rebut their criticism in this Note.
\end{abstract}

\section{The essence of our proof} \label{section-introduction}

\begin{figure}
  \begin{center}
  \includegraphics[width=0.45\textwidth]{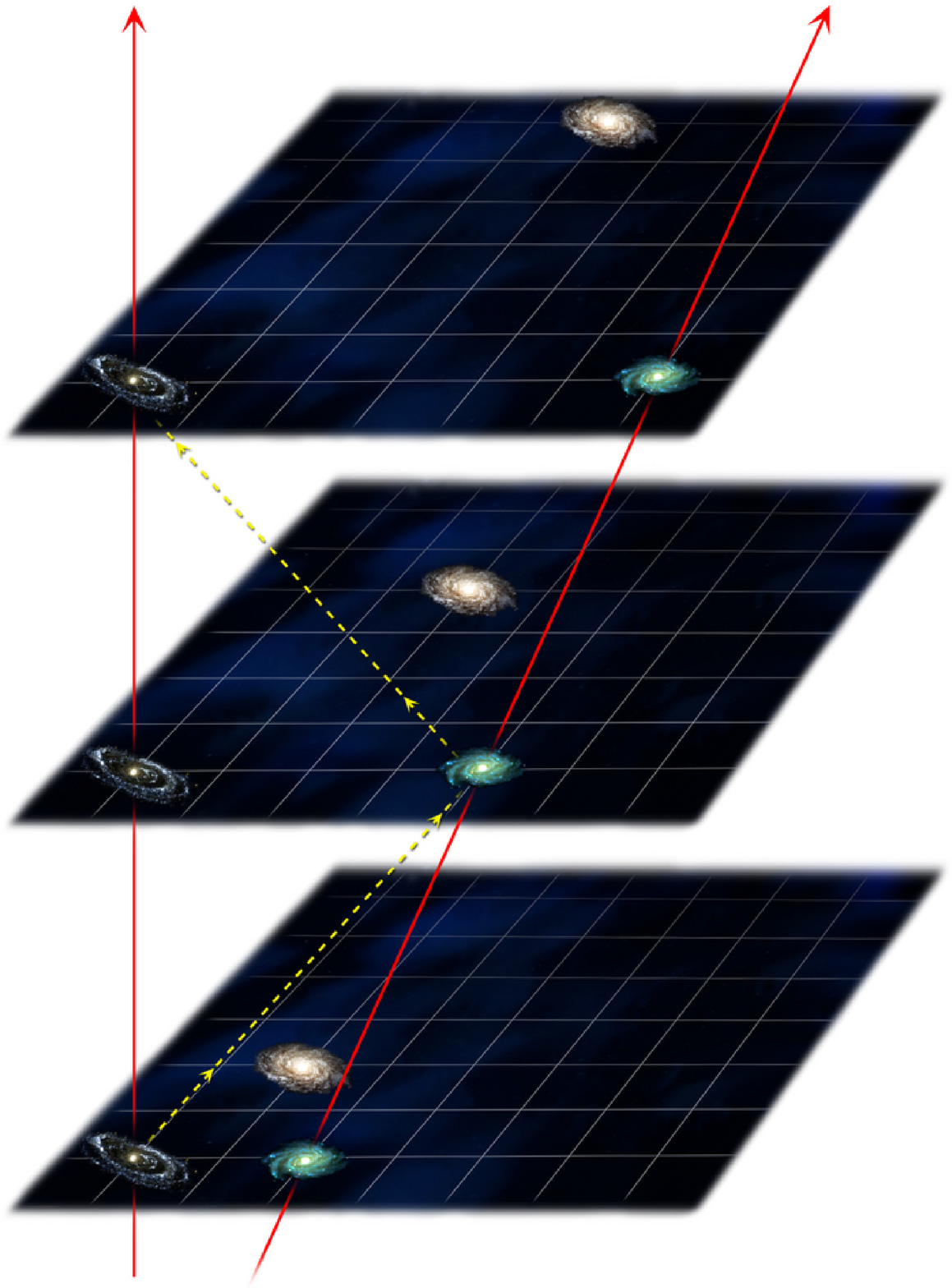}
  \hfill
  \includegraphics[width=0.45\textwidth]{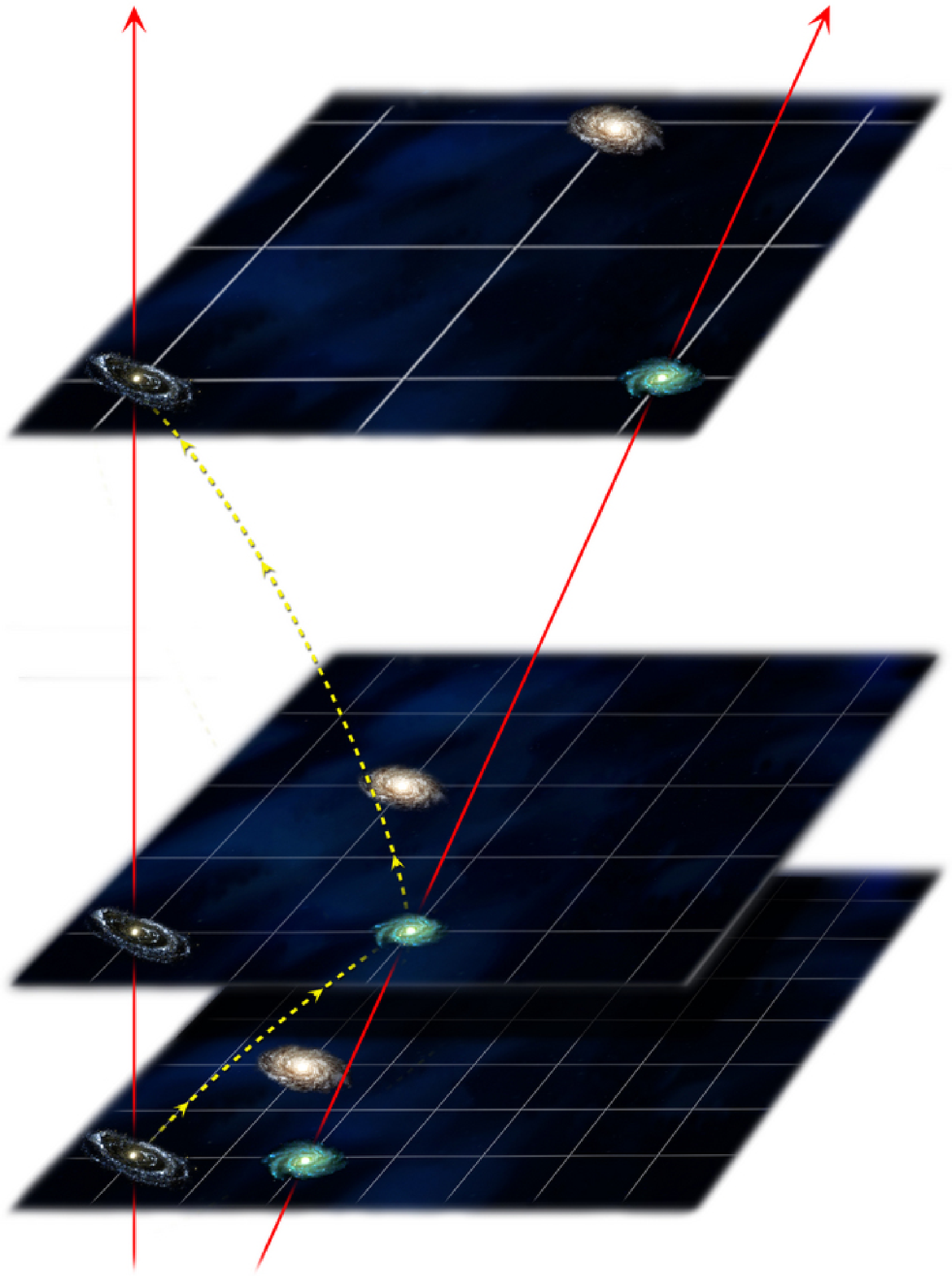}
  \caption{Radar measurement of distance. The light trajectories reveal
  {\it different} spacetime structures in the case of non-expandig
  and expanding space. Obviously, this difference will show up in
  simultaneous measurements of the redshift and distance. Thus measuring the redshift
  and distance allows to establish whether or not space is really expanding.}
  \label{figure}
  \end{center}
\end{figure}

Our claim:

(A) We remark that the concept of ``space'' (in the sense of a
3D submanifold) is not a well-defined ({\it invariant}) concept
in Einstein's general relativity. Therefore the meaning of
``expansion of space'' is indeed ambiguous. However, one
\textit{can} define ``non-expanding space'': as hypersurface in
stationary spacetime, i.e. spacetime equipped with a timelike
Killing vector.

(B) The question of whether the observed cosmological redshift
(given by the Friedmann-Lemaitre-Robertson-Walker metric;
hereafter FLRW) can be distinguished from a purely kinematic
Doppler effect resulting from the motion of particles
(galaxies) in stationary spacetime, has an {\it invariant}
answer. This answer is therefore independent of observers,
coordinates, or conformal factors.

(C) Combined measurements of redshift and radar distance
uniquely distinguish between real cosmological expansion and
the hypothetical motion of galaxies in stationary spacetime.
The reason for this is illustrated in Figure~\ref{figure}.

\section{The criticism and its rebuttal} \label{section-introduction}

The criticism of our arguments is best summarized by Lewis et
al. in \cite{Lewis}. We quote them here in their original
wording, and then present our rebuttal.

\subsection{Rebuttal of the 1st point raised by Lewis et al.}
\label{1st-argument}

``Recently, Abramowicz et al. considered in \cite{eppur} radar
ranging of a distant galaxy in expanding cosmologies and concluded
that the fact that the radar and Hubble distance, from $d =
H_0\,v$, differ in all but an empty universe, that space must
really expand. In a counter argument, \cite{Chodorowski} again
considers radar ranging in open cosmological models. Instead of
examining distances, he focuses upon the transit time of light in
usual cosmological coordinates and its conformal representation.
With this he reveals that in the former coordinates the paths are
asymmetrical in transit time, taking longer on the return journey,
whereas in conformal coordinates, the light travel times to and
from the distant galaxy are equal. Hence, he concludes that the
expansion of space is a coordinate dependent effect which can be
made to disappear with the correct coordinate transform, and
therefore the expansion of space is not a physical phenomenon.''

Rebuttal: The argument by Chodorowski in \cite{Chodorowski}
unfortunately misses the whole point. Our {\it invariant}
statement is based on a gedanken experiment, in which a
fundamental observer measures the interval of his proper time
between two events and a redshift. These measurements are
independent of coordinate or conformal representations.
Chodorowski instead discusses a different problem: of ``the
transit time of light in usual cosmological coordinates and its
conformal representation''. Here the answer is obviously
coordinate--dependent, but the problem is by no means
equivalent to our invariantly defined gedanken experiment.

\subsection{Rebuttal of the 2nd point raised by Lewis et al.}
\label{2nd-argument}

``In their recent work, Abramowicz et al. showed that, in all but
an empty universe, distances derived from the Hubble law and radar
ranging differ and hence "one must conclude that space is
expanding". But how is this difference occurring? Is the expansion
of space acting on a light ray (or even a rocketeer) as they
travel through the universe? We can think of space as a rubber
sheet that stretches to wash out peculiar motions and drives
everything back into the Hubble flow (see \cite{Barnes}). However,
it is the presence of matter that necessitates the inclusion of
gravitational forces upon the motion of the rocketeers and it is
this - the changing gravitational influence of matter in the
universe on the rocketeers - that causes the increasing asymmetry
moving down the panels in Figure \ref{figure}, not that space
physically expands.''

Rebuttal: The measurements of redshift and proper time interval
for receiving the radar echo that we discussed in \cite{eppur}
depend directly on the spacetime geometry. We considered in
\cite{eppur} two reference metrics: the standard FLRW
cosmological metric (\ref{cosmology}) and the Minkowski metric
(\ref{Minkowski}),
\bea ds^2 &=& dt^2 - R^2(t)\left[ dr^2 + r^2 (d\theta^2 +\sin^2\theta
d\phi^2)\right], \label{cosmology} \\
ds^2 &=& dt^2 - \left[ dr^2 + r^2 (d\theta^2 +\sin^2\theta
d\phi^2)\right]. \label{Minkowski} \eea
Because the geometry in (\ref{cosmology}) and (\ref{Minkowski})
is fixed, the measurements that we consider depend neither on
field equations nor on matter distribution.

Ellis and collaborators \cite{Ellis-et-al, Ellis}, playing
devil's advocates, considered a model of a non-expanding
Universe, assuming a static spherically--symmetric (SSS)
metric,
\be ds^2 = G^2(r)\,dt^2 - \left[ dr^2 + F^2(r)(d\theta^2 +\sin^2\theta
d\phi^2)\right]. \label{Ellis} \ee
There is obviously no expansion of space in the SSS Universe.
The observed cosmological redshift-magnitude $(z, m)$ relation
is explained as a purely gravitational effect. The price to pay
is high: not only is the SSS Universe spatially inhomogenous,
but in addition one must assume that we live near its center,
and that at the antipodal location there should be a static
fireball that mimics the Big Bang.

Notice that the gedanken experiment we consider in \cite{eppur}
would immediately rebut the SSS model, since it shows that
radar-measured distances to galaxies do not increase with time.

Some people might perhaps be willing to pay the high price
mentioned above in order to avoid the ``expansion of space''
alternative (the hostility to which eludes our understanding),
and consider the motion of galaxies in the non-expanding space
described by the metric (\ref{Ellis}). Even in this case, a
coordinate--independent {\it experiment} can distinguish
between the true expansion (also of space) that is consistent
with the FLRW metric (\ref{cosmology}) and its kinematic
imitation consistent with the non-expanding space in
(\ref{Ellis}).

Following \cite{Ellis-et-al}, we leave demonstrating this as an
exercise for the reader.

\subsection{Rebuttal of the 3rd point raised by Lewis et al.}
\label{3rd-argument}

``In closing, we state that it is a fools errand to search for the
truth of the existance of expanding space; not only because it is
dependant upon a choice of coordinates, but also because general
relativity is represented by Newtonian physics in the weak field
limit and the global behaviour of the FLRW metric always reduces
to Newtonian gravity in the limit of the local universe with no
need for expanding space. While the expansion of space is a valid
(but dangerous picture when working with the equations of
relativity), any attempts to obtain observations to address the
question of whether galaxies are moving through static space or
are carried away by the expansion of space are doomed to
failure.''

Rebuttal: The above statement that ``general relativity is
represented by Newtonian physics in the weak field limit and
the global behaviour of the FLRW metric always reduces to
Newtonian gravity in the limit of the local universe with no
need for expanding space'' is {\it not} an argument against the
expansion of space. Newton's gravity is a mathematical
approximation to Einstein's theory but is \textit{conceptually}
totally different. Indeed, the consistent and coherent
description of the situation here is this: according to
Einstein's theory, cosmological space expands. As we argued in
\cite{eppur}, the expansion of space implies a non-zero
curvature of spacetime. At small distances, Newtonian theory
should give a correct description of the physical situation.
The only way to describe non-zero curvature in Newton's theory
is through gravitational potential. Thus gravitational
potential should also provide a way of describing the physical
effects of the expansion of space in Newton's theory, with
non--expanding space. Indeed, as explained by Bondi (see e.g.
\cite{Peacock}), to second order, it is indeed correct to think
of the cosmological redshift as a combination of Doppler and
gravitational redshifts in Newton's physics.

\section{Final statement} \label{section-final}

Although the concept of space is not well defined in Einstein's
relativity, one can prove, as we did in \cite{eppur}, that the
statement that the cosmological redshift may be described as a
Doppler effect in non-expanding space is \textit{false}. Eppur si
espande.

\vskip 0.5truecm

Support from Polish Ministry of Science grants NN203 394234 (S.B.)
and N203 009 31/1466 (M.A.A.) is acknowledged.

\end{document}